\documentclass[usenatbib]{mnras}
\usepackage[utf8]{inputenc}
\usepackage{amsmath}
\usepackage{amssymb}
\usepackage{graphicx}
\usepackage[T1]{fontenc}
\usepackage{ae,aecompl}
\usepackage{subfig}

\title[Estimating stellar birth radii]
  {Estimating stellar birth radii and the time evolution\\ of the Milky Way's ISM metallicity gradient}

\author[I.~Minchev, F.~Anders, et al.]
{I.~Minchev,$^{1}$
F.~Anders,$^{1}$
A.~Recio-Blanco,$^{2}$
C.~Chiappini,$^{1}$
P.~de~Laverny,$^{2}$
\newauthor
A.~Queiroz,$^{3.4}$
M.~Steinmetz,$^{1}$ 
V.~Adibekyan,$^5$
I.~Carrillo,$^{1}$
G.~Cescutti,$^{6}$
\newauthor
G.~Guiglion,$^{1}$  
M.~Hayden,$^{7}$
R.~S.~de~Jong,$^{1}$ 
G.~Kordopatis, $^{2}$
S.~R.~Majewski,$^{8}$
\newauthor
M.~Martig,$^9$
B.~X.~Santiago$^{3,4}$
\\
$^{1}$Leibniz-Institut f\"ur  Astrophysik Potsdam (AIP), An der Sternwarte 16, 14482 Potsdam, Germany\\
$^3$Instituto de F\'isica, Universidade Federal do Rio Grande do Sul, Caixa Postal 15051, 91501-970 Porto Alegre, Brazil\\
$^4$Laborat\'orio Interinstitucional de e-Astronomia - LineA, Rua Gal. Jos\'e Cristino 77, 20921-400 Rio de Janeiro, Brazil\\
$^5$Instituto de Astrof\'isica e Ci\^encias do Espa\c{c}o, Universidade do Porto, CAUP, Rua das Estrelas, 4150-762 Porto, Portugal\\
$^{6}$INAF, Osservatorio Astronomico di Trieste, I-34131 Trieste, Italy \\
$^{7}$Sydney Institute for Astronomy, School of Physics, A28, The University of Sydney, Sydney NSW 2006, Australia\\
$^{8}$Department of Astronomy, University of Virginia, Charlottesville, VA 22904-4325, USA\\
$^9$Astrophysics Research Institute, Liverpool John Moores University, 146 Brownlow Hill, Liverpool L3 5RF, UK
}

\pagerange{\pageref{firstpage}--\pageref{lastpage}} \pubyear{2018}

\pubyear{2018}

\begin{document}
\label{firstpage}
\pagerange{\pageref{firstpage}--\pageref{lastpage}}
\maketitle

\begin{abstract}
We present a semi-empirical, largely model-independent approach for estimating Galactic birth radii, $r_{birth}$, for Milky Way disk stars. The technique relies on the justifiable assumption that a negative radial metallicity gradient in the interstellar medium (ISM) existed for most of the disk lifetime. Stars are projected back to their birth positions according to the observationally derived age and [Fe/H] with no kinematical information required. Applying our approach to the AMBRE:HARPS and HARPS-GTO local samples, we show that we can constrain the ISM metallicity evolution with Galactic radius and cosmic time, $\rm[Fe/H]_{ISM}(r, t)$, by requiring a physically meaningful $r_{birth}$ distribution. We find that the data are consistent with an ISM radial metallicity gradient that flattens with time from $\sim-0.15$~dex/kpc at the beginning of disk formation, to its measured present day value ($-0.07$~dex/kpc). We present several chemo-kinematical relations in terms of mono-$r_{birth}$ populations. One remarkable result is that the kinematically hottest stars would have been born locally or in the outer disk, consistent with thick disk formation from the nested flares of mono-age populations and predictions from cosmological simulations. This phenomenon can be also seen in the observed age-velocity dispersion relation, in that its upper boundary is dominated by stars born at larger radii. We also find that the flatness of the local age-metallicity relation (AMR) is the result of the superposition of the AMRs of mono-$r_{birth}$ populations, each with a well-defined negative slope. The solar birth radius is estimated to be $7.3\pm0.6$ kpc, for a current Galactocentric radius of 8~kpc.
\end{abstract}


\section{Introduction}
\label{sec:intro}

It is now well established that stars in galactic disks move away from their birth places, or migrate radially. This has been suspected for a long time for the Milky Way. \cite{grenon72, grenon89} identified an old population of super-metal-rich stars (hereafter SMR), currently found close to the solar radius, but with kinematics and abundances similar to inner disk stars. SMR stars have metallicities that exceed the present-day ISM and that of young stars in the solar vicinity by $\gtrsim 0.25$ dex. The metallicity in the solar vicinity, however, is not expected to have increased much in the last $\sim4$ Gyr because of the low star formation rate (SFR) at the solar radius during this period, possibly combined with continuous infall of primordial gas onto the disk (e.g., \citealt{chiappini03, asplund09}). \cite{sellwood02} interpreted the scatter in the local age-metallicity relation (AMR) as the result of radial migration, where stars below/above the mean arrive from the outer/inner disk. \cite{haywood08} argued that stars with low-$\rm[\alpha/Fe]$ and low-metallicity were born in the outer disk. \cite{minchev14} found a decline in the stellar velocity dispersion as a function of [Mg/Fe] in data from the RAVE \citep{steinmetz06} and SEGUE \citep{yanny09} surveys (seen also in Gaia-ESO data by \citealt{guiglion15, hayden18}), reasoning that these high-[$\alpha$/Fe], kinematically cool stars in each narrow [Fe/H] subpopulation have arrived from the inner disk. SMR stars were concluded to have originated in the inner disk using different data sets (e.g., GCS - \citealt{casagrande11}, FEROS - \citealt{trevisan11}, RAVE - \citealt{kordopatis15, wojno16}). \cite{quillen18} showed that the migration rate in the last $\sim1$~Gyr can be constrained by the spread in metallicity of open cluster currently found close to the solar radius, and thus constrain the strength of the Milky Way spiral structure.

\begin{figure}
\centering
\includegraphics[width=0.8\linewidth, angle=0]{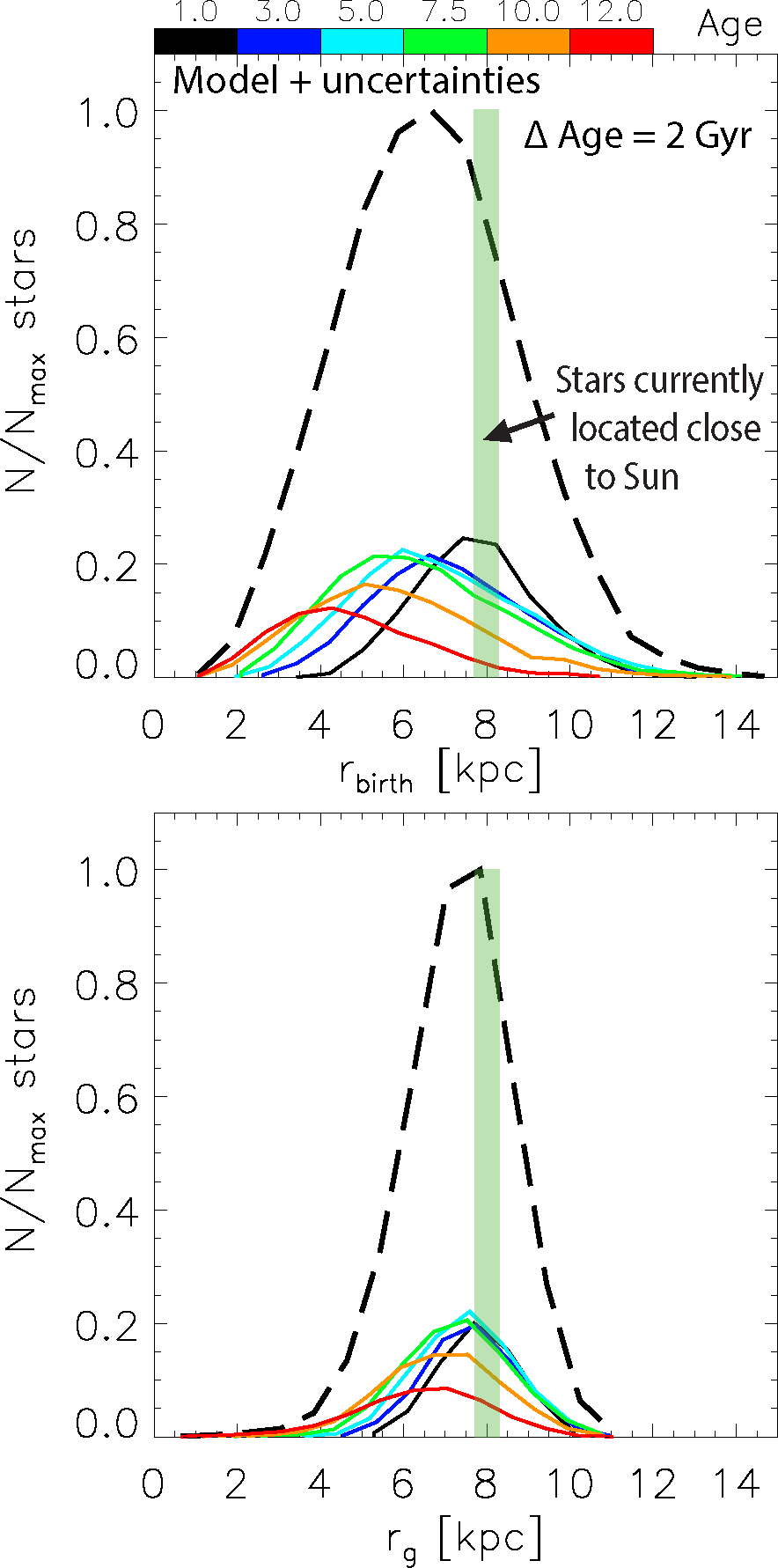}
\caption{
Illustrating the effects of radial migration and blurring. {\bf Top:} the dashed black curves shows the distribution of birth radii, $r_{birth}$, for stars found close to the Sun (green strip) at redshift zero, for the MCM13 model. Also shown are subsamples of stars, grouped by common age (mono-age populations, color curves) with bin width $\rm \Delta age=2$~Gyr and median values as stated in the color bar. The $r_{birth}$ distribution peaks shift to smaller radius the older the population (as color-coded). This results from the inside-out disk formation and the longer exposure to migration mechanisms for older stars.  
{\bf Bottom:} Same as top panel, but showing the stellar guiding radii estimated as $r_g=rv_\phi/V_c$, where $r$ is Galactic radius, $v_\phi$ is the Galactocentric tangential velocity, and $V_c$ is the circular velocity. The overall distribution is much narrower, with mono-age populations having closely spaced peaks but extended tails toward the inner disk, more so for the older population.
}
\label{fig:mcm}
\end{figure}

This picture is also supported theoretically. A number of papers on the topic of stellar radial migration have been produced since the seminal work by \cite{sellwood02}, who proposed that transient spiral arms were responsible (cf. \citealt{roskar08a}). \cite{mq06} identified a new galactic disk heating mechanism, resulting from the interaction among long-lived spiral modes moving at different pattern speeds and suggested that this also creates radial mixing, a phenomenon explored later in detail by \cite{mf10} and \cite{minchev12a}. \cite{grand12a} showed that the spirals seen in disk morphology (as opposed to the power spectrum) in their N-body simulations were moving close to corotation at all radii, which was interpreted by \cite{comparetta12} as the effect of long-lived multiple modes \citep{tagger87, quillen11}. Finally, external perturbations from infalling satellites can also create radial mixing \citep{quillen09,bird13}, mostly in the outer disk, but also at smaller radii because of the triggered increase in bar and spiral structure amplitudes. Using their hybrid chemo-dynamical model, \cite{mcm13} (hereafter, MCM13) showed that SMR stars in their simulated solar neighborhood originated almost exclusively from the radial range $3<r<5$ kpc (their Fig.~3). This origin coincides with the end of the Galactic bar, which is expected to produce most of the radial migration in the inner disk after its formation (e.g., \citealt{minchev12a, dimatteo13}). 

The effect of radial migration on a solar neighborhood-like simulated disk volume, resulting from the MCM13 model, is shown in the top panel of Fig.~\ref{fig:mcm}. Stars found at 8 kpc at the final time can originate from a wide range of birth radii, $r_{birth}$. The $r_{birth}$ distribution peak shifts to smaller radius for older mono-age populations (stars in narrow age bins, as color-coded). This results from the inside-out disk formation and the longer exposure to migration mechanisms for older stars. In the bottom panel of Fig.~\ref{fig:mcm} we show the guiding radius, $r_g$, distribution of these stars. Stars away from 8~kpc are found in the solar vicinity close to their apo- and pericenters with older mono-age populations having more extended tails to smaller radii. This effect, referred to as "blurring", is sometimes invoked to explain the scatter in the AMR, arguing that no radial migration is necessary (e.g., \citealt{haywood16}). It is obvious here, however, that the stars causing blurring have migrated themselves.

Knowledge about the birth radii of stars has important implications also for the field of exoplanet research. Stars hosting planets show the tendency to have smaller guiding radius, suggesting smaller birth radii \citep{haywood09,adibekyan14}. \cite{haywood09} proposed that the giant planet formation efficiency is not directly linked to the stellar metallicity, as typically accepted, but on a parameter related to the distance from the Galactic center, and thus likely, the birth radius.

From the above discussion it is evident that accounting for the phenomenon of radial migration is very important for advancing our understanding of the Milky Way formation and evolution. The question then arises: can we develop a method for recovering the birth positions of stars we observe today without invoking ever more complex modeling?

One way to do this is by considering the evolution of chemical abundance gradients, for example that of the metallicity, [Fe/H]. In a pioneering paper, \cite{matteucci89} established that the observed negative Milky Way disk metallicity gradient results from an inside-out disk formation. That the Milky Way disk has formed from the inside out is now evident from the more centrally concentrated older stellar populations (e.g., \citealt{bensby11, bovy12a}). There has been, however, little progress on understanding the time evolution of the ISM abundance gradient - a key ingredient for recovering the disk star formation history (SFH) and, thus, understanding the assembly history of the Galaxy. Classical chemical evolution models in the past have been typically constrained by the present-day ISM abundances, the shape of the metallicity distribution function, super novae rates, gas and stellar densities. Assumptions about gas infall and gas flows determine the SFR with cosmic time and radius. This freedom has led to degeneracies among models in the path taken to reproduce the observations today. For example, depending on the choice of initial conditions, the ISM metallicity gradient flattens with time in \cite{prantzos95}, \cite{pilkington12}, and in the pure thin-disk model of \cite{chiappini09a} (used in MCM13), while it steepens with time in \cite{chiappini01} and \cite{cescutti07}, mostly due to a pre-enrichment floor set by the first infall.

An inversion of the metallicity gradient from negative to positive (and in [$\alpha$/Fe] gradient from positive to negative) for samples at increasingly larger distance above the disk midplane, $|z|$, has been observed in most large Milky Way surveys (e.g., SEGUE - \citealt{cheng12a}, RAVE - \citealt{boeche13b}, APOGEE - \citealt{anders14}, Gaia-ESO - \citealt{recio-blanco14}). This inversion was explained by \cite{mcm14} (hereafter, MCM14, see also \citealt{rahimi14, miranda16, ma17}) as the effect of inside-out disk formation and disk flaring of mono-age populations (their Fig.~10). It was shown that gradient inversion can result from the variation with radius of the mixture of stars of different ages: at large $|z|$ stars that are old, metal-poor, and  [$\alpha$/Fe]-rich dominate in the inner disk (due to the inside-out formation), while younger, more metal-rich, and more [$\alpha$/Fe]-poor stars are more abundant in the outer disk (present at high-$|z|$ thanks to disk flaring); this naturally results in a positive metallicity (and a negative [$\alpha$/Fe]) gradient at high $|z|$. Investigating further, \cite{minchev15} showed in two suites of simulations that disk formation in the cosmological context inevitably results in flared mono-age populations, which are more extended the younger the sample because of the inside-out formation. These "nested flares" constitute a geometric thick disk, which does not flare, or in which the flaring is strongly reduced, consistent with the lack of disk flaring found in the observations of external galaxies \citep{vanderkruit82, degrijs98, comeron11}. This model of thick disk formation predicts a negative radial age gradient in geometrically-defined thick disks, which was found to be indeed the case for the Milky Way using APOGEE data \citep{martig16b}.

\begin{figure}
\centering
\includegraphics[width=0.7\linewidth, angle=0]{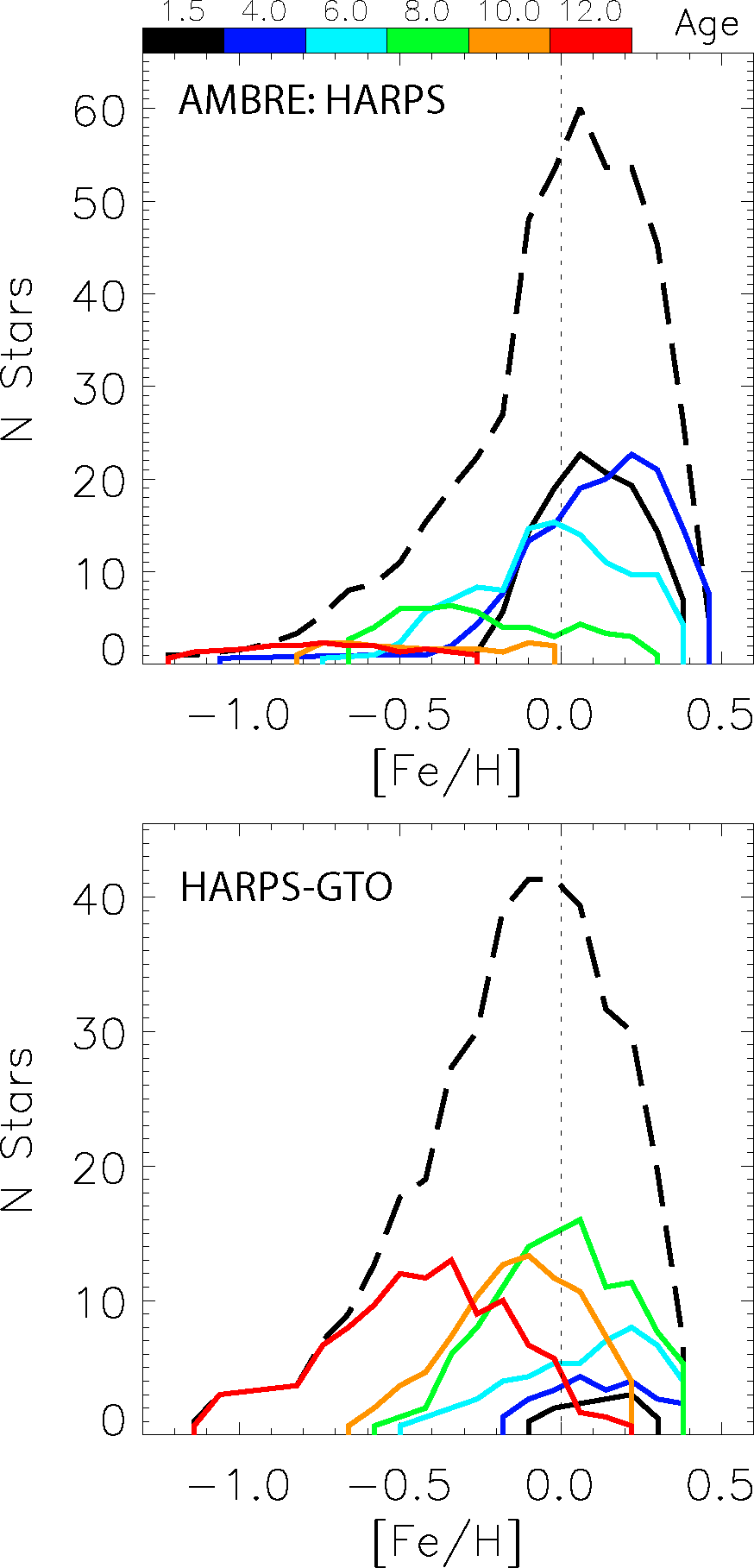}
\caption{  
{\bf Top panel:} The MDF of the total AMBRE:HARPS sample (dashed black curve). Also shown are subsamples of mono-age populations (color curves) using a bin of $\rm \Delta age=2$~Gyr and a median stated in the color bar. 
{\bf Bottom panel:} Same but for the HARPS-GTO sample. For both samples the youngest stars are concentrated around [Fe/H]=0 dex, the metal-rich tail is composed of intermediate age stars, and the oldest population is found in the metal-poor tail.
}
\label{fig:mdf}
\end{figure}

\begin{figure*}
\centering
\includegraphics[width=1.\linewidth, angle=0]{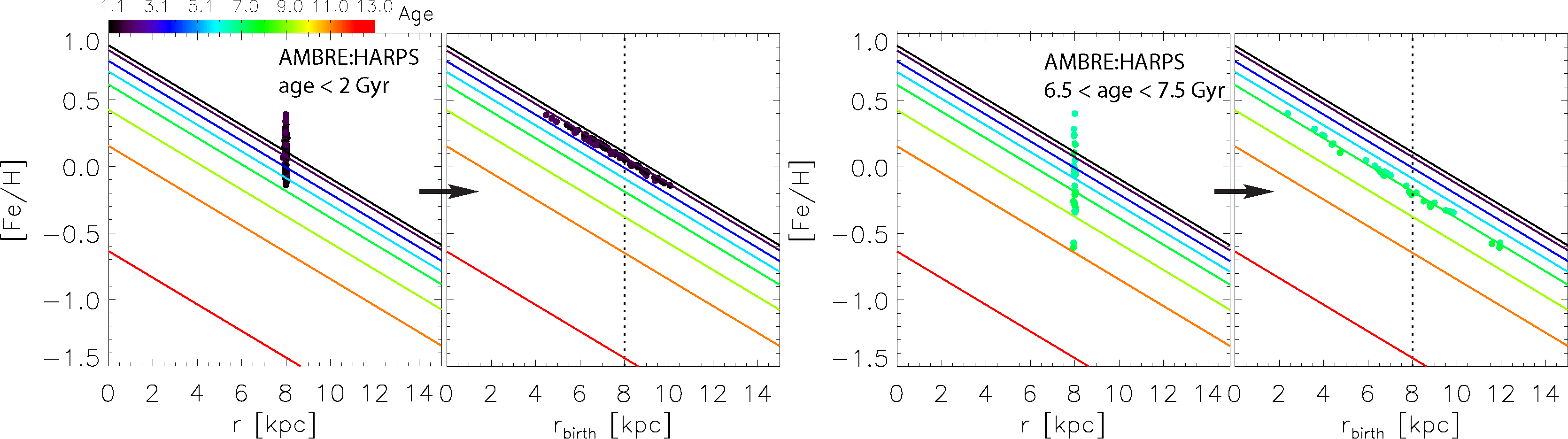}
\caption{
Illustration of our birth radius estimation method for two mono-age groups from the AMBRE:HARPS sample, as indicated. The color-coded lines represent an arbitrary [Fe/H] ISM gradient evolution with time. The scatter in [Fe/H] of stars with the same age (leftmost and third panels) is interpreted as the result of stars born at different Galactic radii. We can estimate the birth radii by projecting stars in radius along the corresponding age gradient, as shown in the second and fourth panels. 
}
\label{fig:ex}
\end{figure*}

The idea presented by MCM14 was that, while metallicity gradients of mono-age populations are always negative, flattening or even an inversion can result due to the uneven mixture of different ages with radius. \cite{anders17b} recently measured the metallicity gradients of mono-age stellar populations using the high-quality CoRoGEE dataset \citep{anders17a} - a sample of red giants with chemical abundances measured by APOGEE \citep{majewski10,majewski17} and asteroseismic ages from the CoRoT satellite. It was indeed found that the metallicity gradient is negative for all ages, in the range $6<r<15$~kpc and $|z|<0.3$~kpc, although the total population shows inversion at high $|z|$ (e.g., \citealt{anders14}). This example highlights the importance of having good stellar ages to facilitate the correct interpretation of chemo-kinematical relations.

In this work we develop a largely model-independent method for estimating stellar birth radii of any stellar sample with good abundance measurements and age estimates. As a byproduct we provide, for the first time, a constraint on the ISM abundance gradient evolution with time. Because no kinematical information is used in our procedure, we can later include the stellar kinematics for unbiased analyses.

\section{Data}
	
We consider two local high-quality datasets, mostly composed of main sequence trunoff and subgiant stars, both using spectra taken by the HARPS instrument. Stellar parameters in each case, however, are derived differently and ages are estimated using different codes, as described below. Both samples have been well described and analyzed in the literature. With this work we aim to provide additional information, the stellar birth radii, that will help with the further scientific interpretation of these data.

\subsection{AMBRE:HARPS}

The AMBRE project is a uniform analysis of archival ESO spectra, as described in \cite{delaverny13}. Spectra are taken from the HARPS instrument, complemented by the TGAS catalog. Stellar parameters are derived using the MATISSE algorithm \citep{recio-blanco06}, as described in \cite{depascale14}. Typical errors are $\rm T_{eff}<100$~K and $\sim0.05$~dex in [Fe/H] and [Mg/Fe] \citep{mikolaitis17}. Most stars lie within 50 pc from the Sun, with fractional parallax error of $<5\%$. To estimate Galactocentric velocities, distances and proper motions are used from Gaia DR1/TGAS \citep{gaia16}, along with radial velocities determined by AMBRE.

Ages are estimated for main sequence turnoff and subgiant stars using isochrone fitting, as described by Hayden et al. (submitted). The AMBRE:HARPS sample we use in this work consists of 488 stars, identical to the one presented by \cite{hayden17} except for 6 stars. Those were discarded because they were on radial orbits and therefore may have not been born in the disk. Our method in general would also not work for accreted populations, of which only a small fraction, if any, may be present in our very nearby samples.

\begin{figure*}
\centering
\includegraphics[width=1.\linewidth, angle=0]{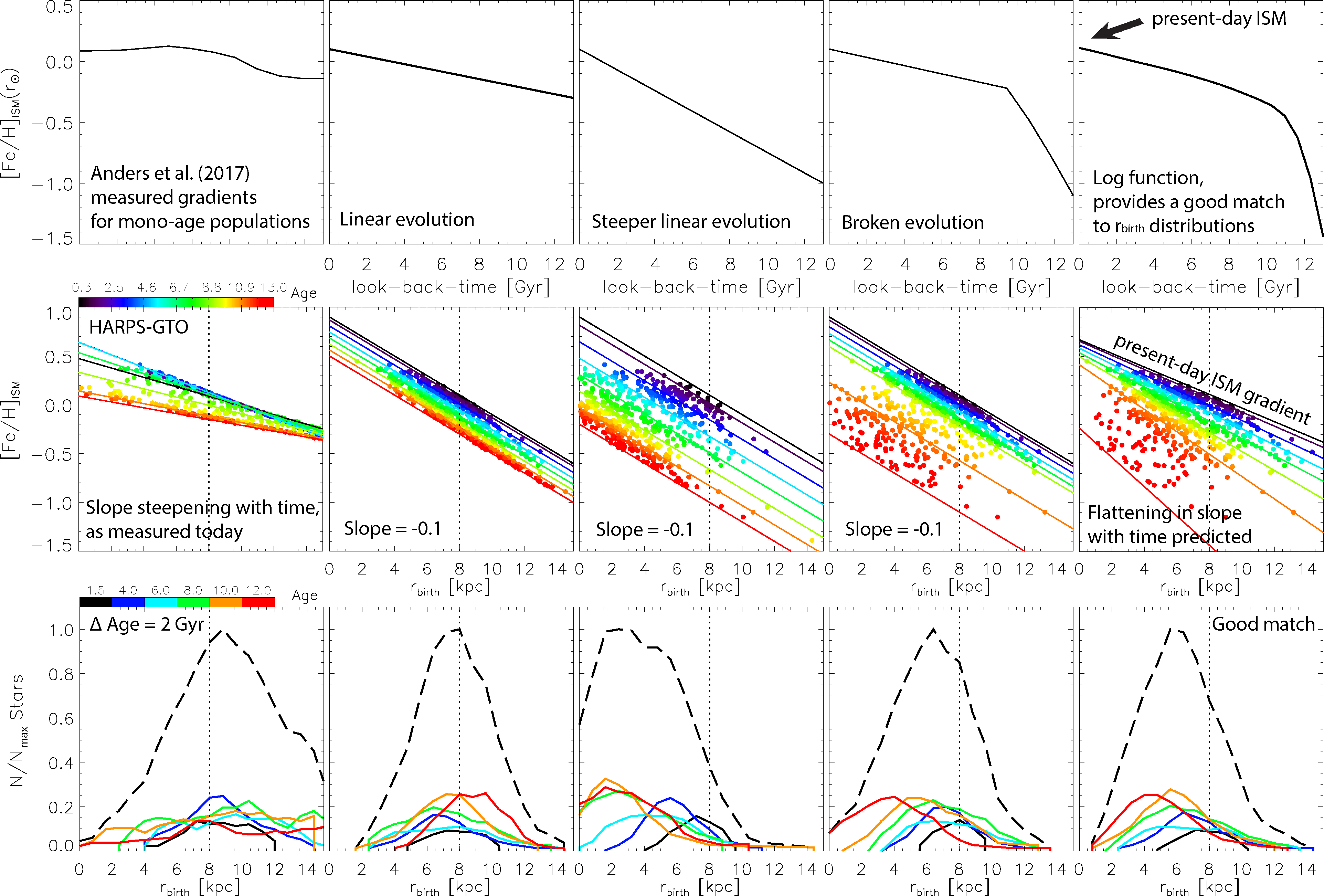}
\caption{
Exploring the effect of different possibilities for the ISM metallicity variation with radius and time, $\rm [Fe/H]_{ISM}(r,t)$, using our HARPS-GTO sample. In the leftmost column we use the metallicity variation with age at the solar radius, $\rm [Fe/H](r_\odot, t)$ (top panel) and the slope variation with age (for uniformly spaced times, middle panel) provided by Anders et al. (2017a) to estimate $r_{birth}$, making the unjustified assumption that age here corresponds to look-back-time. The $r_{birth}$ distributions we derive (bottom-leftmost panel), using our method illustrated in Fig.~\ref{fig:ex}, are clearly unphysical for this model, as the majority of stars appear to have been born far in the outer disk, and look very different from what we expect from a typical Galactic disk model (Fig.~\ref{fig:mcm}). This confirms the expectation that the abundance gradient variation with age is not the same as the time evolution. We next consider different possibilities for $\rm [Fe/H]_{ISM}(r,t)$, while keeping the slope fixed at $-0.1$~dex/kpc and varying the metallicity time evolution at the solar radius $\rm [Fe/H]_{ISM}(r_\odot, t)$. A shallow linear increase with time (second column) suggests that older stars were born farther out in the disk than younger ones, in conflict with an inside-out disk formation. A steeper linear evolution (third column) results in $r_{birth}<0$ kpc for $\sim25\%$ of stars. When we combine an initially steep increase in [Fe/H] with a slower one for the past $9-10$~Gyr (fourth column), birth radii distributions start to look as expected, in particular the decrease of peak for older ages in the expected radial range. Exploring the time variation of the slope, while being constrained by the present-day one, we find that a good match is a radial ISM metallicity gradient at early times of $-0.15$~dex/kpc, flattening toward redshift zero (rightmost column).
}
\label{fig:der}
\end{figure*}

\subsection{HARPS-GTO}

The second sample we use is the HARPS-GTO data, studied extensively in a number of works (e.g., \citealt{adibekyan11, adibekyan12, haywood13, anders14, delgadomena17}). We make use the stellar parameters and chemical abundances from a recent reanalysis of the HARPS-GTO sample \citep{delgadomena17}, which has a median signal-to-noise ratio of 240, $\rm T_{eff}$ uncertainty of 30 K, [Fe/H] uncertainty of 0.02 dex, and [Mg/Fe] uncertainty of 0.04 dex. Surface gravities were estimated from Hipparcos \citep{vanleeuwen07}. Proper motions come from Gaia DR1/TGAS when available, or Hipparcos otherwise. Radial velocities were estimated from HARPS spectra \citep{adibekyan12}.
Ages and distances were estimated with the StarHorse code \citep{santiago16, queiroz18}, as described in \cite{anders18}. 
We imposed the following quality criteria on ages and abundances: $\rm\delta[Mg/Fe]<0.07$~dex, $\rm\delta age/age<0.25$ or $\rm\delta age<1$~Gyr, where $\rm\delta age$ and $\rm\delta[Mg/Fe]$ are the corresponding uncertainties. The resulting sample has 603 stars. We make the additional quality cut $\rm 5300<T_{eff} < 6000$ K when presenting the AMR, the age-[Fe/H] relation, and the [Mg/Fe]-[Fe/H] plane (see \S\ref{sec:amr}, \S\ref{sec:age}, and \S\ref{sec:afe}, respectively), ending up with 388 stars. The overlap between the two samples, after the above quality cuts, is 239 and 152 stars, respectively. 

For both samples Galactocentric velocities are estimated adopting a solar radius of $r_\odot=8$~kpc and Local Standard of Rest of $V=220$~km/s.

Since the procedure for deriving ages is different for each dataset (although both rely on isochrone fitting), we normalize the two to the range ${\rm min(age)\lesssim age\lesssim13}$~Gyr. This compresses the two samples from their maximum ages of 15 Gyr (AMBRE:HARPS) and 13.7 Gyr (HARPS-GTO). 

The metallicity distribution function (MDF) of each sample is shown in Fig.~\ref{fig:mdf}. The MDFs of mono-age subpopulations are also plotted, as color-coded. Both total-sample MDFs peak at $\rm [Fe/H]\approx0$~dex and both have similar ranges. The lighter metal-poor tail in AMBRE:HARPS is caused by the lower number of old stars (see green, orange, and red curves) compared to HARPS-GTO.

\section{Method}
\label{sec:method}

Our method of stellar birth radius estimation is illustrated in Fig.~\ref{fig:ex}. The leftmost and third panels show the AMBRE:HARPS [Fe/H] distribution for two mono-age populations, as indicated by the dots, color-coded by age. Let us assume for now that we know the time evolution of the ISM metallicity gradient, indicated by the color lines of slope $\rm d[Fe/H]_{ISM}/dr=-0.1$~dex/kpc. To find the birth radius, $r_{birth}$, we simply project stars to the metallicity gradient corresponding to their age, as shown in the second and fourth panels. Note that the younger age bin has less scatter in [Fe/H] compared to the older mono-age population, as expected (e.g., \citealt{anders17b}), which naturally results in a narrower range of birth radii.

If we knew the evolution of the ISM metallicity with radius and time, $\rm [Fe/H]_{ISM}(r,t)$, then we could use this simple method to find the birth places of all stars in our samples. The problem is that this function is unknown. We next show that we do not need prior knowledge of $\rm [Fe/H]_{ISM}(r,t)$. We will use the following constraints to estimate both the birth radii and ISM metallicity variation with time directly from the data:

\begin{figure*}
\centering
\includegraphics[width=0.9\linewidth, angle=0]{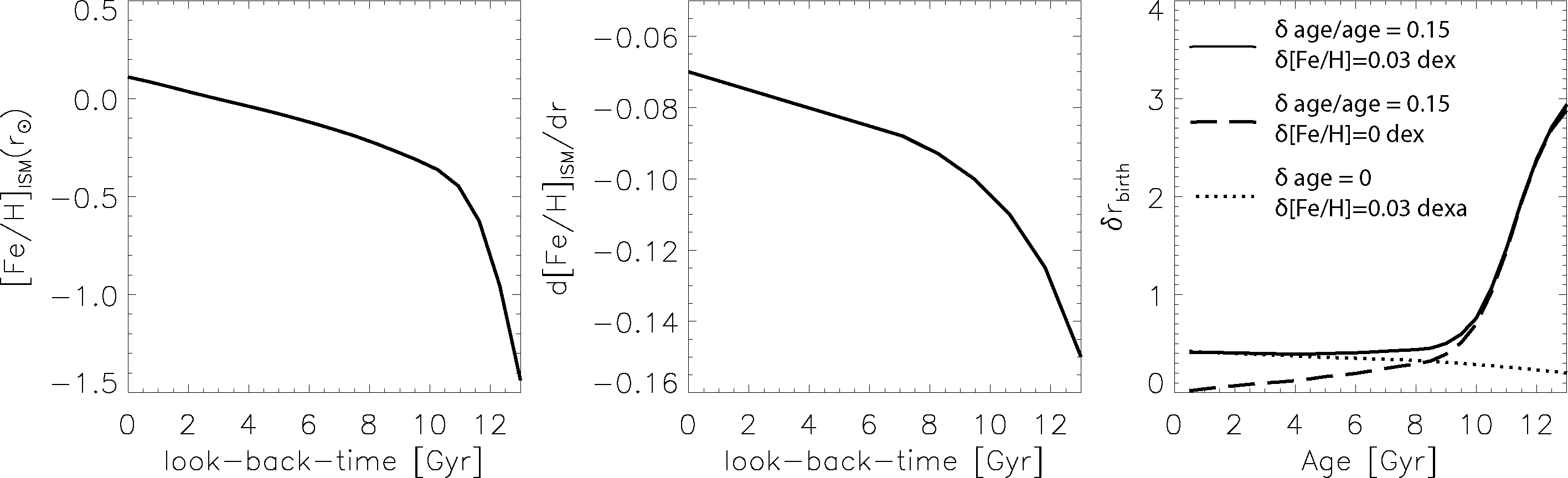}
\caption{
ISM metallicity variation with time at the solar radius, $\rm [Fe/H](r_\odot, t)$, (left) and the ISM metallicity slope variation with time, $\rm d[Fe/H]_{ISM}/dr$, (middle), resulting from the $r_{birth}$ distribution match to Fig.~\ref{fig:mcm}, shown in the bottom rightmost panel of Fig.~\ref{fig:der}. The right panel shows the $r_{birth}$ uncertainty as a function of stellar age, estimated from the HARPS-GTO [Fe/H] and age typical uncertainties: $\rm \delta[Fe/H] = 0.03$~dex and $\rm \delta age\sim15\%$ (solid curve). An increase is seen at earlier times due to the fast chemical enrichment at the onset of disk formation. Setting $\rm \delta[Fe/H] = 0$ (dashed curve) or $\rm \delta age=0$ (dotted curve) shows that the metallicity uncertainty dominates the $r_{birth}$ error at $\rm age\lesssim8$~Gyr,  while $\rm \delta age$ dominates for older stars. 
}
\label{fig:err}
\end{figure*}

\begin{itemize}
	\item {\it Present-day ISM metallicity gradient of $-0.07$~dex/kpc} - this is slightly steeper than observations of classical Cepheids ($\sim100$~Myr old, e.g., \citealt{genovali14}, $\sim-0.06$~dex/kpc);
	\item {\it ISM metallicity gradient always negative} (with the possible exception for $\rm age > 10$~Gyr) - must be true, since it is also observed for mono-age groups today (CoRoGEE sample - \citealt{anders17b}, see also \S\ref{sec:grad}); dynamically it is not easy to invert gradients, except possibly in the disk outskirts \citep{minchev11a};
	\item {\it Resulting $r_{birth}$ distribution of youngest mono-age bin peaks at the solar radius} - not much migration is expected for stars younger than $\sim1$~Gyr, therefore, the majority of these must be locally born. This is also supported by the fact that the gradient of CoRoGEE stars with $\rm age<1$~Gyr lies on top of the Cepheid gradient \citep{anders17b}, which traces the past $\sim100$~Myr;
	\item {\it Resulting $r_{birth}$ distributions of mono-age populations peak at smaller radius the older the population} - more migration is expected for older populations and the disk forms inside-out. This is consistent with the MCM13 model (see Fig.~\ref{fig:mcm}) and expected from a range of disk formation simulations (e.g., \citealt{roskar08a, brook12}, \citealt{bird13, ma17});
	\item {\it Resulting $r_{birth}$ distribution of oldest mono-age bin peaks at $\sim4$ kpc} - if this peak was located at a smaller radius, negative birth radii would result for a sizable fraction of stars (see Fig.~\ref{fig:der}). The generality of this result needs to be checked in other simulations;
	\item {\it Stars are born with the same metallicity at a fixed radius and cosmic time, i.e., the ISM is well-mixed at a given radius} - this appears to be the case for redshift zero \citep{nieva12}, but this assumption may not hold for earlier times. Possible problems with this assumption are discussed in \S\ref{sec:err}.
		
\end{itemize}

In Fig.~\ref{fig:der} we explore the variation of birth radius distributions for different $\rm [Fe/H]_{ISM}(r,t)$ possibilities.  What if the gradients we observe today for mono-age populations represented the ISM gradients, which would be the case if stars did not migrate? In the leftmost column of Fig.~\ref{fig:der} we use the slopes and metallicity variation with time at the solar radius, $\rm [Fe/H](r_\odot, t)$, provided by \cite{anders17b} to estimate $r_{birth}$ for the HARPS-GTO sample, making the unjustified assumption that age here corresponds to look-back-time. The $r_{birth}$ distributions we derive (bottom-leftmost panel) are clearly unphysical, as the majority of stars appear to have been born far in the outer disk and look nothing like the expectation from a typical galactic disk model (Fig.~\ref{fig:mcm}, top). The [Fe/H] gradients of mono-age populations would represent the ISM metallicity evolution only if there were no migration. 
The bottom-leftmost panel of Fig.~\ref{fig:der} shows that this cannot be the case. It is expected that stellar gradients flatten with time and that the ISM gradient at any time must have been steeper than the corresponding mono-age population measured today (see discussion in \S\ref{sec:flattening}).

We next consider different possibilities for $\rm [Fe/H]_{ISM}(r,t)$, while keeping the slope fixed at $-0.1$~dex/kpc and varying the metallicity evolution at the solar radius $\rm [Fe/H]_{ISM}(r_\odot, t)$. A shallow linear evolution (Fig.~\ref{fig:der}, second column) suggests that older stars were born farther out in the disk than younger ones, in conflict with an inside-out disk formation. A steeper linear evolution (third column) results in $r_{birth}<0$ kpc for $\sim30\%$ of stars. When we combine an initially steep increase in [Fe/H] with a slower one for the past $9-10$~Gyr (fourth column), birth radii distributions start to look like our requirements (see above), in particular the decrease of peak for older ages in the expected radial range. 

This exercise shows that the time evolution of [Fe/H] is mostly responsible for the $r_{birth}$ peak of mono-age distributions. Exploring different slopes, while being constrained by the present-day one, we find that a good match is a radial ISM metallicity variation at early times of $-0.15$~dex/kpc, flattening toward redshift zero (rightmost column of Fig.~\ref{fig:der}). Both the $\rm [Fe/H]_{ISM}(r_\odot)$ time evolution and the slope time evolution were found to be well described by log functions, shown in Fig.~\ref{fig:err}. 

Propagating the typical uncertainties in the HARPS-GTO data of $\rm \delta[Fe/H] = 0.03$~dex and $\rm \delta age\sim15\%$, we estimate $\delta r_{birth}\lesssim0.5$~kpc for stars of $\rm age\lesssim8$~Gyr, increasing for older populations due to the fast chemical enrichment at high redshift (see Fig.~\ref{fig:err}, right panel). Setting $\rm \delta[Fe/H] = 0$ (dashed curve) or $\rm \delta age=0$ (dotted curve) shows that the metallicity uncertainty dominates the $r_{birth}$ error at $\rm age\lesssim8$~Gyr, while $\rm \delta age$ dominates for older stars. Further discussion on the $r_{birth}$ error can be found in \S\ref{sec:err}.

\begin{figure*}
\centering
\includegraphics[width=0.9\linewidth, angle=0]{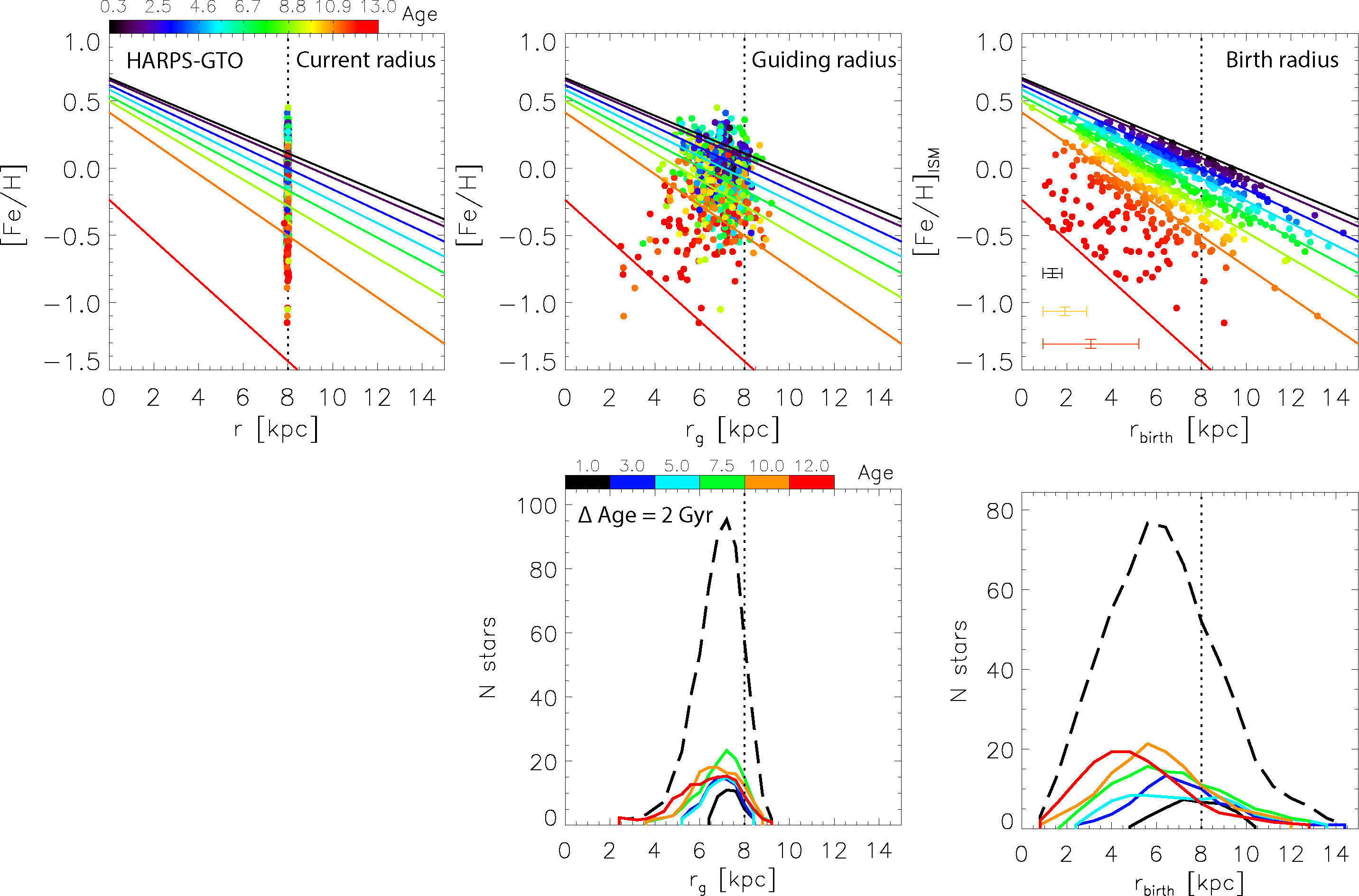}
\caption{
{\bf Left:} [Fe/H] vs Galactic disk radius, $r$, showing the currently observed positions of HARPS-GTO stars, color-coded by age. Our match to the ISM [Fe/H] gradient evolution is shown by the lines of negative slope for eight uniformly spaced times.
{\bf Middle column:} [Fe/H] vs guiding radius, $r_g$ (top) and the corresponding distributions of mono-age populations (bottom). 
{\bf Right column:} as in middle, but for the estimated birth radius, $r_{birth}$, identical to the rightmost middle and bottom panels in Fig.~\ref{fig:der}. Typical error bars for young, intermediate, and old stars are shown in the upper panel, estimated for $r_{birth}$ as described in \S\ref{sec:method}. Just as in the model shown in Fig.~\ref{fig:mcm}, the $r_g$ mono-age distributions are weighted toward the inner disk, more so for older populations, but the peaks are clustered just inside $r=8$~kpc.
}
\label{fig:rg}
\end{figure*}

\section{Results}

The results from the above simple experiment and a handful of high-quality data can already address important Galaxy evolution questions in a completely new and much more intuitive way than previously possible.

\subsection{Guiding vs birth radii}

The top row of Fig.~\ref{fig:rg} shows [Fe/H] vs observed radius, $r$ (left), guiding radius, $r_g$ (middle), and birth radius, $r_{birth}$ (right) for the HARPS-GTO stars. The bottom panels show the corresponding distributions of mono-age populations. Our match to the ISM [Fe/H] gradient evolution is shown for eight uniformly spaced times. The right column is identical to the rightmost middle and bottom panels of Fig.~\ref{fig:der}. We remind the reader that only age and [Fe/H] were used for obtaining $r_{birth}$. The guiding radius distribution is strongly weighted toward the inner disk, more so for older mono-age populations. The latter is indicative of inside-out disk formation, justifying our requirement on the shape of $r_{birth}$ distributions (see \S\ref{sec:method}). 

The total and mono-age population distributions of $r_g$ resulting from the data are remarkably similar to the model shown in Fig.~\ref{fig:mcm} in the peak separation, width, and skewness. It should be kept in mind that the guiding radii from data and model are completely independent.

\begin{figure*}
\centering
\includegraphics[width=0.8\linewidth, angle=0]{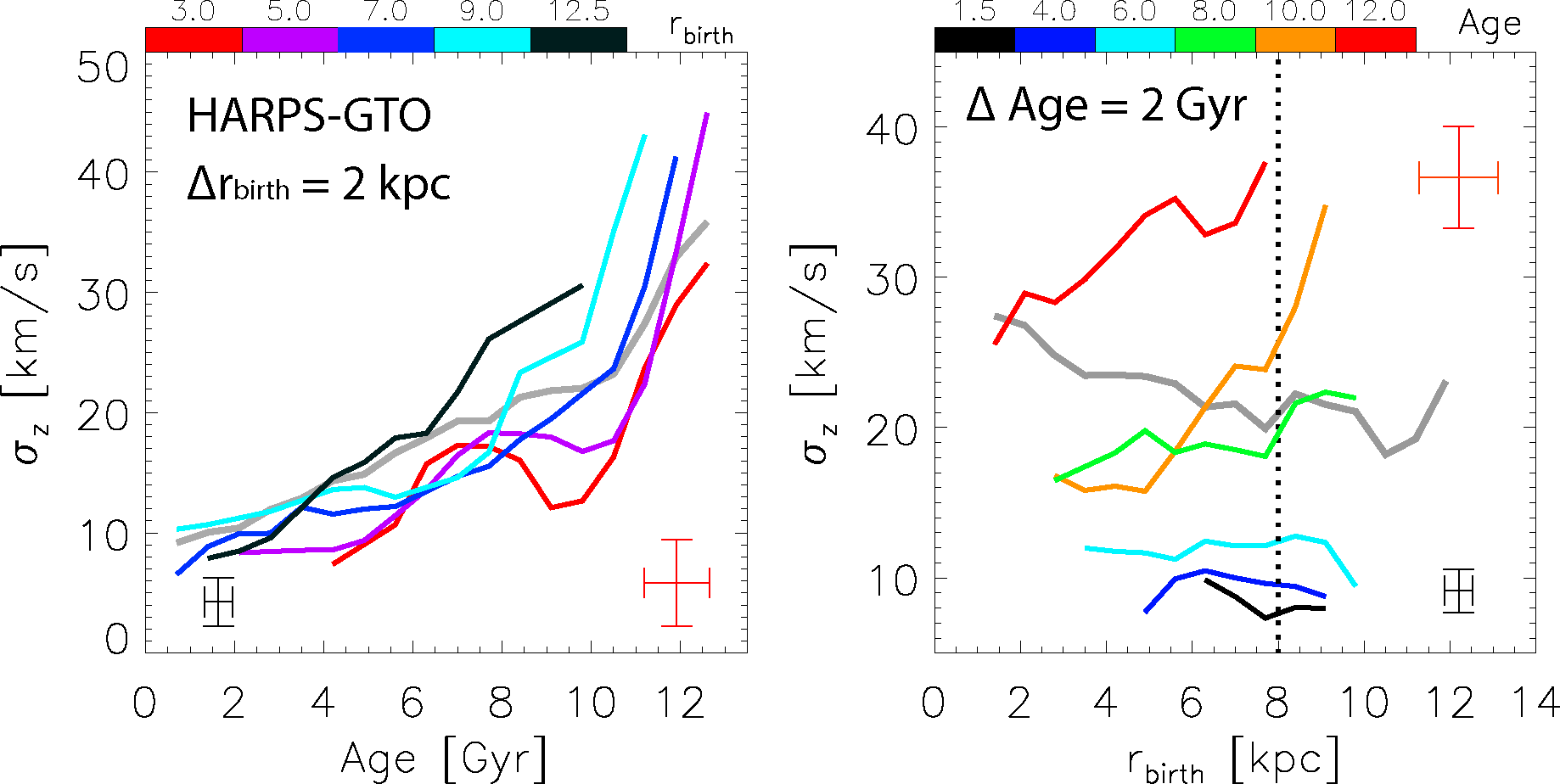}
\caption{
{\bf Left:} The age-$\sigma_z$ (vertical velocity dispersion) relation for our total HARPS-GTO sample (gray curve) and for mono-$r_{birth}$ populations (color-coded). Typical error bars are shown, estimated for $\sigma_z$ as the two standard deviations of 1000 realizations in a bootstrapping calculation and for $r_{birth}$ as $\left<r_{birth}\right>/\sqrt{N}$ (and similarly for age), for N=8 stars in a bin. Stars born at larger radii tend to be kinematically hotter for the same age, especially at $\rm age\gtrsim8$~Gyr. A step (or a jump) in $\sigma_z$ at $\sim10$~Gyr is found in the total sample, but much better defined for individual mono-$r_{birth}$ radii, excluding the outermost bin. 
{\bf Right:} Vertical velocity dispersion, $\sigma_z$, variation with $r_{birth}$. Although an overall increase of $\sigma_z$ is seen with decreasing birth radius in the total sample (gray curve), a positive relation is present for older mono-age populations (as color-coded), i.e., the hottest stars near the Sun today were born locally or in the outer disk. This positive slope flattens toward redshift zero, eventually turning slightly negative for age $\lesssim4$~Gyr. This plot, resulting from observations, is remarkably similar to expectations from chemo-dynamical modeling (see Fig.~5 by Minchev et al. 2014).
}
\label{fig:avr}
\end{figure*}

\subsection{Vertical age-velocity dispersion relation (AVR) and thick disk formation}

In the left panel of Fig.~\ref{fig:avr} we show the relationship between age and stellar vertical velocity dispersion, $\sigma_z$, for all stars (gray curve) and for mono-$r_{birth}$ populations (as color-coded) using the HARPS-GTO sample. The velocity dispersion is estimated as the standard deviation of stars in each age bin, constrained to a minimum of 8 stars per bin. We also show typical error bars, corresponding to two standard deviations of 1000 realizations in a bootstrapping calculation.

Examining the vertical AVR we find that stars born at the same time but at larger radii, tend to be kinematically hotter, especially at $\rm age\gtrsim8$~Gyr. A step in $\sigma_z$ at $\rm age\sim10$~Gyr is found in the total sample (gray curve), which is much better defined for mono-$r_{birth}$ populations, excluding the outermost $r_{birth}$ bin. The latter is not surprising, as the disk most likely did not extend beyond 12 kpc 10~Gyr ago. For $\rm age\gtrsim10$~Gyr, stars born inside 4 kpc (red curve) have about 1/3 the velocity dispersion of those born beyond 10 kpc (black curve). This result also explains the trend seen in Fig.~4 by \cite{hayden17}, where AMBRE:HARPS data were used to show that the low-[Mg/Fe] sequence has higher vertical AVR compared to the high-[Mg/Fe] sequence, especially at $\rm age<8$~Gyr. 

Such a step in the AVR has been previously found observationally \citep{quillen01,freeman02} and suggested to result from dynamical heating due to the last massive merger, in agreement with simulations (e.g., \citealt{martig14b}). Alternatively, using the APOSTLE cosmological simulations, \cite{navarro18} concluded that the AVR is a reflection of the gradual thinning of the disk, rather than the effect of dynamical processes. MCM13 argued that it was a combination of both stars born hot at high redshift \citep{bournaud09,forbes12} and subsequent heating from mergers \citep{brook04} that created the kinematically defined thick disk. 

In the right panel of Fig.~\ref{fig:avr} we show how $\sigma_z$ varies with $r_{birth}$, which is another way to display the information contained in the AVR of mono-$r_{birth}$ populations. Although an overall increase is seen with decreasing birth radius (gray curve), a positive relation is present for older mono-age populations (as color-coded), i.e., the kinematically hottest stars near the Sun today were born locally or in the outer disk, as already inferred from the AVR. This positive slope flattens toward redshift zero, becoming slightly negative for age $\lesssim4$~Gyr. The trends in this figure are remarkably similar to the MCM13 model expectation shown in Fig.~5 of \cite{minchev14}. 

The result that stars born at larger radii are kinematically hotter than those born at smaller radii is remarkable as it indicates that outward migration cools the local disk, rather than heat it, as has been proposed in the past \citep{schonrich09b, roskar13}. This finding supports work showing that migration does not contribute to thick disk formation (e.g., \citealt{minchev12b}, MCM14, \citealt{vera-ciro14, grand16}, see also discussion in \S\ref{sec:intro}). 

\begin{figure*}
\centering
\includegraphics[width=1.\linewidth, angle=0]{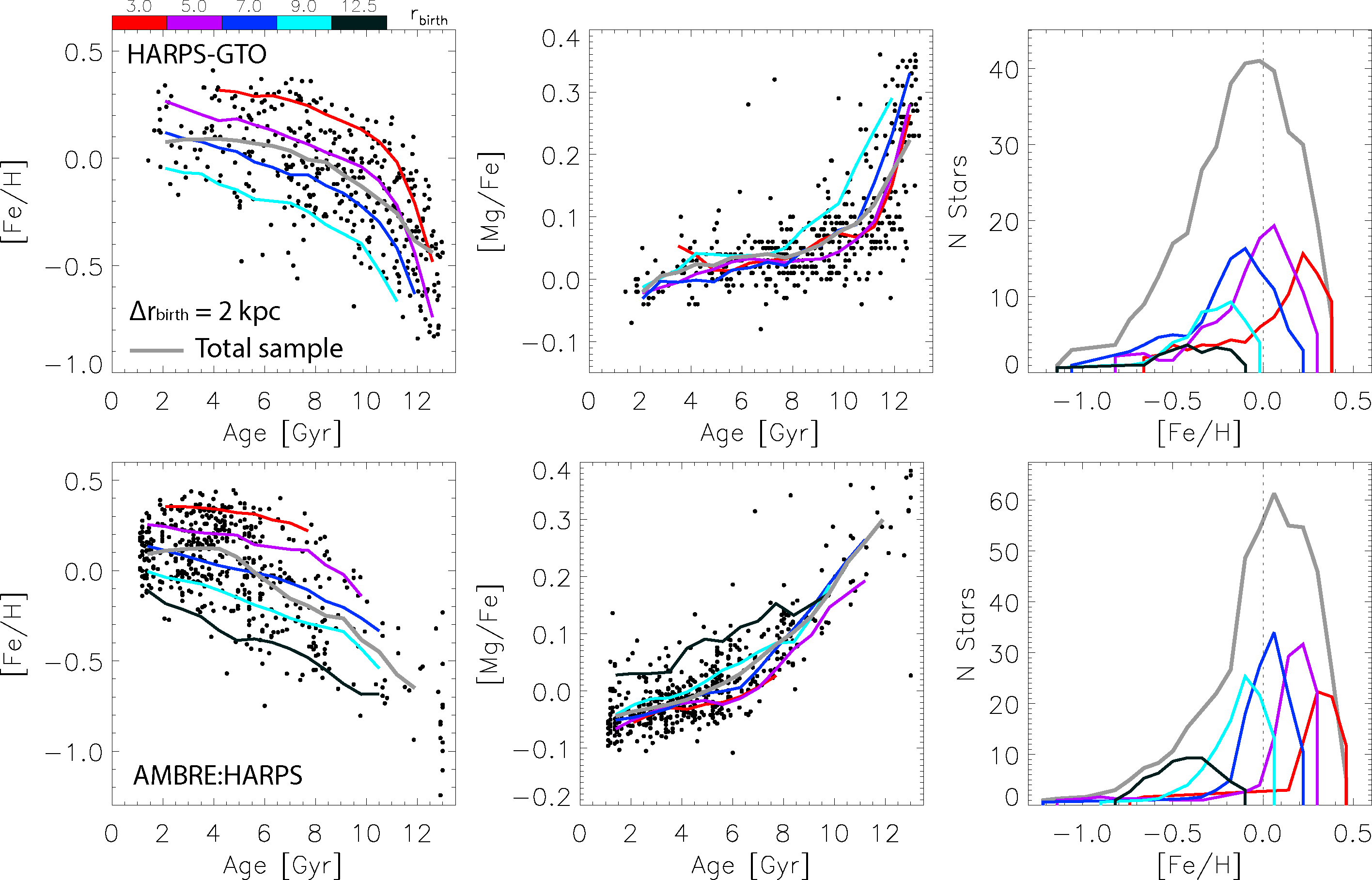}
\caption{
{\bf Left:} The local age-metallicity relation (AMR) using HARPS-GTO (top) and AMBRE:HARPS data (bottom). The gray curves show the total mean of each sample. The color-coded lines represent five mono-$r_{birth}$ populations, with median values indicated in the color bar and a bin width $\Delta r_{birth}=2$~kpc. Well-defined AMRs exist for stars with common birth radii. The AMR is shifted downward for outer radii, which is a result of the negative ISM radial metallicity gradient. Note that there is no galactic model involved here. {\bf Middle:} Same as left row but for the [Mg/Fe]-age relation. {\bf Right:} The MDF of the total samples (gray curves) with mono-$r_{birth}$ subpopulations color coded, as in the left and middle columns. Mono-$r_{birth}$ populations appear to have self-similar shapes (reverse log-normal distribution) with a cutoff at higher values, the lower the birth radius.
}
\label{fig:amr}
\end{figure*}

\subsection{Age-metallicity relation (AMR)}
\label{sec:amr}

In the left column of Fig.~\ref{fig:amr} we show the AMR for HARPS-GTO (top) and AMBRE:HARPS (bottom) data. The gray curves show the total mean of each sample. The color-coded lines represent five mono-$r_{birth}$ populations, with median values indicated in the color bar and a bin width $\Delta r_{birth}=2$~kpc. This is not an unbiased result since the $r_{birth}$ estimate is based on age and [Fe/H]. Nevertheless, this figure reflects how the functional form of the ISM metallicity gradients determines the shape of the mono-$r_{birth}$ AMRs.

We can see that well-defined AMRs with slopes steeper than that of the total HARPS-GTO data exist for stars with common birth radii.\footnote{The outermost $r_{birth}$ bin is missing for the HARPS-GTO data, due to the low number of stars.} This may be the explanation for the puzzling flatness of the local AMR found for the past $\sim10-11$ Gyr (e.g., \citealt{edvardsson93, lin18}). Similar trends are seen for AMBRE:HARPS, except for the decline in the mean at $\rm age\gtrsim5$~Gyr, which is due to the low number statistics for old ages.

The AMR is shifted downward for outer radii, which is a result of the negative ISM radial metallicity gradient. These trends are similar to expectations from chemical evolution models (e.g., \citealt{chiappini09a}, MCM13 Fig.~4), but note that there is no Galactic model involved here, just the requirement to keep $r_{birth}$ distributions of mono-age populations physically meaningful (as outlined at the beginning of \S\ref{sec:method}). Similar and independent results are obtained by \cite{anders18} and Chiappini et al. 2018 (in prep) by using the t-SNE dimensionality-reduction technique \citep{vandermaatel08}, applied to chemical abundance space.

\subsection{Age-[Mg/Fe] relation}
\label{sec:age}

The middle column of Fig.~\ref{fig:amr} is similar to the left one, but showing the [Mg/Fe]-age relation. A change from a slow to a fast increase of [Mg/Fe] with age is seen at $\sim10$~Gyr (for HARPS-GTO) and $\sim7$~Gyr (for AMBRE:HARPS), similar to previous works (e.g., \citealt{anguiano12, haywood13, bensby14, feuillet18}). This transition is related to the fast chemical enrichment at high redshift, when the high-[Mg/Fe] stars are formed during a period with high SFR (e.g., \citealt{chiappini97, fuhrmann11, haywood16}). 

The slow-to-fast-increase transition becomes less prominent (and possibly shifts to younger ages) for outer radii (cyan and black curves), consistent with an early strong SFR in the inner disk, related to the inside-out disk formation.

\begin{figure*}
\centering
\includegraphics[width=0.9\linewidth, angle=0]{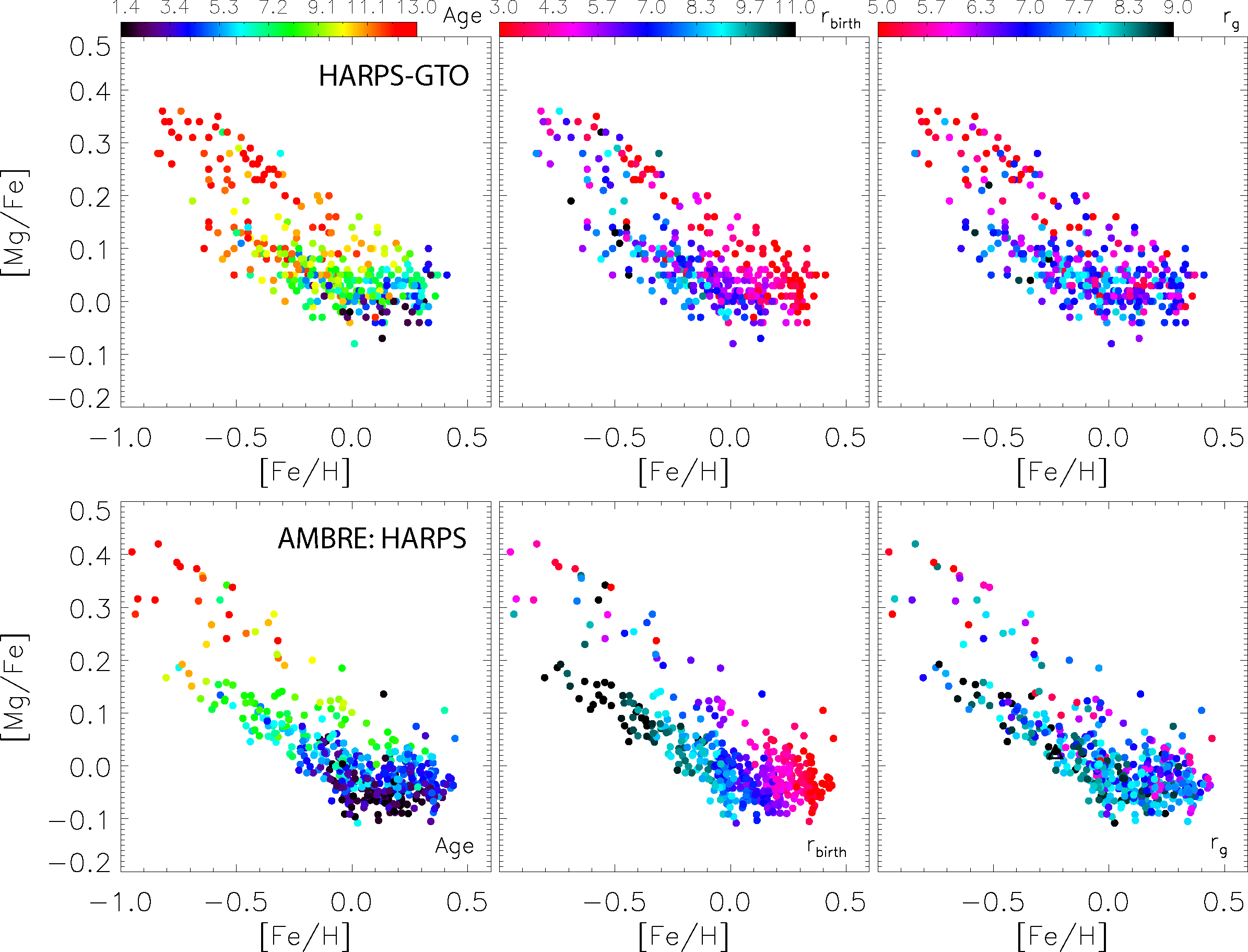}
\caption{
[Mg/Fe]-[Fe/H] plane for the HARPS-GTO sample color-coded by age (left), birth radius (middle) and guiding radius (right). Old stars populate the high-[Mg/Fe] sequence and some of the metal-poor end of the low-[Mg/Fe] sequence. Stars with the smallest birth radii are those with the highest [Fe/H] values. Stars with the largest birth radii belong to the metal-poor end of the low-[Mg/Fe] sequence.
}
\label{fig:afe}
\end{figure*}

\subsection{Metallicity distribution function (MDF)}

In the right column of Fig.~\ref{fig:amr} we show the MDFs for our two samples (gray curves), binned by stars with similar birth radii (as color-coded). Stars beyond $\rm [Fe/H]\gtrsim0.2$ dex have birth radii inside 5~kpc, which supports the prediction by the MCM13 chemo-dynamical model (their Fig.~3, see also estimate by \citealt{kordopatis15} using RAVE data). This figure also looks very similar to the bottom-right panel of Fig~15 by \cite{feuillet18}, resulting from their chemical evolution model taking radial migration into account. 

Mono-$r_{birth}$ populations appear to have self-similar shapes (reverse log-normal distribution) with a cutoff at higher values, the lower the birth radius. 

\subsection{[Mg/Fe]-[Fe/H] plane}
\label{sec:afe}

Fig.~\ref{fig:afe} shows the [Mg/Fe]-[Fe/H] plane for our two samples, color-coded by age (left), birth radius (middle) and guiding radius (right). The color scales for the middle and right columns are slightly compressed on each side to show better contrast. The actual ranges of the $r_g$ and $r_{birth}$ histograms can be seen in the bottom row of Fig.~\ref{fig:rg}.

For both data sets the high-[Mg/Fe] sequence starts very old at the low-[Fe/H] end, becoming younger with increasing metallicity. For HARPS-GTO, some significantly old stars are also present in the low-[Mg/Fe] sequence at $\rm [Fe/H]<0$~dex. SMR stars ($\rm [Fe/H]>0.25$) have the smallest birth radii (Fig.~\ref{fig:afe}, middle), as was already seen in the MDFs of mono-$r_{birth}$ populations (right panel of Fig.~\ref{fig:amr}). The high-[Mg/Fe] metal-poor stars have a range of birth radii from the smallest up to about solar radius. Stars with the largest birth radii belong to the metal-poor end of the low-[Mg/Fe] sequence, as has been expected from their angular momenta (or high rotational velocities, e.g., \citealt{haywood08}). This is especially prominent for the AMBRE:HARPS data.

The high-[Mg/Fe] metal-poor stars show a mixture of guiding radii (right panel of Fig.~\ref{fig:afe}), similar to the mixture of birth radii we find.\footnote{Note that the color bars in the middle and right panels show different scales.} It can be seen from this figure that the strongest changes in angular momentum (or migration, which is the difference between birth and guiding radius) has occurred for the SMR stars, most of which have guiding radii in the range $7-8$~kpc (cyan color in right panel of Fig.~\ref{fig:afe}). That these stars were born in the very inner disk was also concluded by \cite{anders18} and Chiappini et al. 2018 (in preparation), using also the HARPS-GTO sample but a completely different approach.

\subsection{Solar birth radius}

Due to its metallicity and age, the Sun has been suggested to originate from inside the solar circle. For example, \cite{wielen96} estimated a value of $6.6\pm0.9$ kpc and MCM13 found $4.6 < r_{birth} < 7.6$~kpc. Using $\rm [Fe/H]=0\pm0.05$~dex \citep{asplund09} and $\rm age=4.56\pm0.11$~Gyr \citep{bonanno02}, we estimated $r_{\odot, birth}=7.3\pm0.6$~kpc. 

Setting $\rm \delta age=0$ does not affect the $r_{\odot, birth}$ uncertainty and setting $\rm \delta [Fe/H]=0$ we find $\delta r_{birth, \odot}=0.1$~kpc, indicating that the metallicity uncertainty dominates for this particular age (see right panel of Fig.~\ref{fig:err}). This estimate is somewhat larger than those quoted above and suggesting that the Sun could have been born very close to its present-day radius, for our adopted $\rm r_\odot=8$~kpc. 

\section{Discussion}

\subsection{Stellar metallicity gradient flattening with time}
\label{sec:flattening}

Because of the inside-out disk formation and the stellar surface density exponential falloff with radius, the metallicity gradient of a mono-age population outside the bulge/bar will always tend to flatten with time due to both radial migration and blurring \citep{roskar08a, minchev11a}. Since older samples have experienced more migration, their gradients are expected to have flattened more, although it should be kept in mind that older stars are also kinematically hot and, thus, not so prone to migration \citep{minchev12a, daniel18}. In any case, the expectation is that the observed gradients of mono-age populations today do not directly represent the evolution of the ISM metallicity, but are likely flatter than they were at their birth time. 

There are two important points in comparing our results with the observed gradient variation with age \citep{anders17b} that suggest our estimate for the ISM gradient evolution is reasonable:

(1) Our estimate for the ISM gradient evolution is consistently steeper than the gradient of mono-age populations, for the same age. This is expected as dynamical processes always tend to flatten the gradient of a narrow age sample as discussed above (see Fig.~5 in MCM13 and Fig.~6 in \citealt{kubryk15b}).

(2) Our estimate for the solar ISM metallicity evolution with time, $\rm [Fe/H]_{ISM}(r_\odot, t)$, is systematically lower than  that of mono-age populations (compare leftmost and rightmost top panels in Fig.~\ref{fig:der}). This is also expected, as flattening of a mono-age stellar gradient happens with a pivot point near the bar corotation (or $\sim1.5$ scale-lengths), resulting in a mean decrease with respect to the birth ISM inside the bulge and a mean increase at the solar radius (see Fig.~5 in MCM13). Scatter around the mean [Fe/H], however, is present for all ages and radii, and is increasing for older populations \citep{anders17b}.

\subsection{Constraints from different Galactic radii and different elements}
\label{sec:difr}

Using only a small patch in the Galactic disk (the HARPS data at $d<100$~pc), the best we could do was estimate the $\rm [Fe/H]_{ISM}(r, t)$ as a linear function in radius. It is clear, however, that the larger the disk area studied, the more powerful this method becomes. The availability of high-quality data from other Galactic radii will allow us to determine the precise functional behavior of the $\rm [Fe/H]_{ISM}(r, t)$ radial profile, and thus improve our $r_{birth}$ estimates. Such datasets will become available from the high-resolution spectroscopic surveys GALAH \citep{desilva15} and Gaia-ESO \citep{gilmore12} within 2-3 kpc from the Sun, and APOGEE, SDSS-V's Milky Way Mapper (MWM, \citealt{kollmeier17}), WEAVE \citep{dalton12}, and 4MOST \citep{dejong12} up to $\sim6-7$ kpc from the Sun, combined with Gaia astrometry and asteroseismic ages from K2 \citep{howell14}, TESS \citep{ricker15}, and PLATO \citep{rauer14}.

Although in this work we only considered [Fe/H], up to $\sim30$ different elements (from, e.g., GALAH, APOGEE, Gaia-ESO, MWM, 4MOST) are now becoming available. Requiring that all elements match the same birth radius distributions of mono-age populations via MCMC modeling will provide additional constraints, allowing for this method to become largely model-independent. 

\subsection{ISM metallicity slope variation with time}
\label{sec:grad}

The steep ISM metallicity gradient we find at high redshift, flattening toward redshift zero, is consistent with cosmological simulations \citep{roskar08a, pilkington12, gibson13, vincenzo18}, hybrid models (MCM13, \citealt{kubryk15b}),  and classical chemical evolution models \citep{prantzos95, chiappini09a, grisoni18}. Although the MCM13 model provided a good match to the results of \cite{anders17b}, the final gradients were systematically lower than those observed, by up to 25\% (see Table A.1. by \citealt{anders17b}). This is in agreement with the model ISM [Fe/H] assigned at birth (from \citealt{chiappini09a}) being flatter than what we find here, especially inside 5 kpc at early times - compare Fig.~5 by MCM13 with the gradient evolution shown in current Fig.~\ref{fig:rg}. 

A flattening with radius inside the bulge region most likely exists for the last couple of Gyr, constrained by the maximum $\rm [Fe/H]\sim0.6$~dex measured in the disk, inferred by the fact that most high-resolution spectroscopic surveys do not find stars beyond $\rm[Fe/H]=0.5-0.6$~dex (e.g., \citealt{rojas-arriagada14, garcia-perez18}). This upper limit, however, is still under debate. The medium-resolution detection of stars with potentially higher metallicities by \cite{do15} and \cite{feldmeier-krause17} was recently followed up with high-resolution spectroscopy by \cite{rich17} who found maximum metallicities of [Fe/H] $\sim0.6$ for those same stars.

\subsection{Flat metallicity gradient at high redshift?}

We found that the early ISM metallicity gradient must have been about twice as steep as it is today, in order to keep the $r_{birth}$ distribution of the oldest stars physically meaningful (see \S\ref{sec:method}). This result, however, may still be degenerate. Depending on the initial conditions, the Galactic disk could have started forming with a flat radial [Fe/H] gradient \citep{chiappini01, cescutti07}. The strong SFR, and thus [Fe/H] increase, in the inner disk at high redshift is then expected to quickly ($\sim1-2$~Gyr) result in a negative metallicity gradient, converging to a model which starts with a negative gradient (as we predict here). If the formation of the Galactic disk did start with a flat gradient, then our method would fail for the oldest stars ($\gtrsim10$~Gyr).

It must be true that up to $\sim10$~Gyr ago the ISM metallicity gradient was negative, given that it is found to be negative for mono-age populations today (see discussion in \S\ref{sec:grad}). For the oldest stars, however, it is not easy to differentiate between an initially flat and a negative ISM metallicity gradient from observations today, because (1) the oldest stars have the largest age uncertainties and (2) gradients of old stars flatten strongly due to migration and blurring. \cite{anders17b} found a gradient of $-0.03\pm0.03$ dex/kpc for the oldest mono-age bin ($\sim12$~Gyr). Improved ages for old stars from asteroseismolgy missions in the near future (see \citealt{miglio17}) will help differentiate between the two scenarios, by considering smaller age bins than we can afford at this time.

\subsection{Dilution of the [Fe/H]$\rm_{ISM}$ with time}

Using the very local HARPS data, we found that the ISM metallicity at all radii always increases with time, which is consistent with present-day's measurements of mono-age populations \citep{anders17b} in the range $6 < r < 15$~kpc. It is possible, however, that at certain times throughout the disk evolution, the ISM was diluted at a given radius (or all radii) from infalling gas or a gas-rich merger. This may have resulted in an overall decrease of the ISM metallicity gradient for a given age group, which is not obvious today because of dynamical effects and an insufficient amount of data. With the availability of large samples covering larger disk area, such an effect can be detected using our methodology applied on smaller age bins.

\subsection{Sources of uncertainty in the $r_{birth}$ estimate}
\label{sec:err}

A star with an overestimated [Fe/H] by $+0.05$~dex will be considered an outward migrator (coming from the inner disk). For a metallicity slope of $0.1$~dex/kpc, the uncertainty introduced in the estimated $r_{birth}$ will therefore be 0.5 kpc (see Fig.~\ref{fig:err}, right panel). There will be, however, a similar fraction of stars with underestimated [Fe/H] and those will be treated as inward migrators. Due to this symmetry, the overall distributions of a mono-age population and, thus, the results presented in Figures \ref{fig:amr}-\ref{fig:afe}, should not be affected much by this problem. The age uncertainties will have an effect similar to that of the metallicity. Since the time evolution of the gradient at early times is very fast, the oldest stars are affected the most, as seen in the right panel of Fig.~\ref{fig:err}. 

One of the assumptions we made in this work was that the ISM has always been well-mixed at a given radius. This appears to be the case at redshift zero from studies of local early B-type stars \citep{nieva12}. Studies of HII regions, however, both in the Milky Way \citep{balser15} and in external galaxies \citep{sanchez15, sanchez-menguiano18}, find azimuthal variations in [O/H] at a given radius, with a scatter on the order of $\sim1$~dex. This may result from spiral structure induced fluctuation in the gas density, which cause deviations of $\sim0.05$~dex (similar to the uncertainty, Spitoni et al. 2018, in preparation), but larger effects can come from gas infall at higher redshift, gas-rich mergers, and radial gas flows. 

Because this effect is largely symmetric around the mean, although it will affect the birth radii of individual stars, we expect that to first order it should not bias significantly distributions, similar to the effect of age and abundance uncertainties. Future work should investigate if and how we can correct for this problem. 

\section{Conclusions}

We presented here a semi-empirical, largely model-independent approach for the estimation of stellar birth radii of Galactic disk stars, $r_{birth}$, using only chemical information and age estimates. The only model dependence was in the age determination and in the requirement to keep $r_{birth}$ distributions of mono-age populations physically meaningful. No galactic disk model or kinematical information were needed. 
An important assumption we made was that a negative radial metallicity gradient in the ISM existed for most of the disk lifetime, supported by the existence of negative metallicity gradients for all mono-age populations today \citep{anders17b}. This was used to project stars back to their birth positions according to the observationally derived age and [Fe/H].

This new technique transforms the stellar age and chemical information into a much more tractable quantity, the birth radius, allowing to present results in an intuitive and easy to interpret manner. The power of this new method is evident in the predictions we could do with just a few hundred stars located very close to the Sun. As a byproduct of the birth radius estimation, we managed to constrain the ISM metallicity as a function of Galactic radius and time. This was done directly from the data for the first time.

We summarize our results as follows:

(1) Applying our approach to the AMBRE:HARPS and HARPS-GTO local samples, we found that the ISM metallicity gradient flattens with time from $\sim-0.15$~dex/kpc for the oldest ages to the observed value today. We showed that the slope of the ISM metallicity controls the peak position of mono-age $r_{birth}$ distributions, while the [Fe/H] evolution at the solar position affects the $r_{birth}$ spread. This allowed us to constrain the ISM metallicity evolution with Galactic radius and cosmic time, $\rm[Fe/H]_{ISM}(r, t)$, simply by requiring realistic $r_{birth}$ distributions of mono-age populations. The latter should be seen only as a weak dynamical constraint, resulting from any inside-out disk formation model. 

(2) We found that the variation of vertical velocity dispersion, $\sigma_z$, with $r_{birth}$ for mono-age populations has a positive slope for old stars, turning flat or slightly negative for the youngest stars. In other words, the deeper in the inner disk old stars come from, the cooler kinematically they are today. This model-independent result argues against earlier claims that the Galactic thick disk was created by radial migration. It appears that in the Milky Way radial migration cools the disk, as argued from a theoretical point of view by MCM13, MCM14, \cite{minchev14}, \cite{ma17}, and \cite{grand16}. The agreement between data and simulations on the $\sigma_z$-$r_{birth}$ relation (compare Fig.~\ref{fig:avr}, right to Fig.~5 by \citealt{minchev14}) is truly remarkable: for the data the figure is produced by going from the final time (today) back to the birth radius, while in the model we go forward in time using an unconstrained simulation in the cosmological context. There is no reason to expect that these will behave similarly and the fact that they do must mean that the simulation has chemo-dynamics similar to the Milky Way. The MCM13 ISM metallicity gradient, however, is not as steep at high redshift as we find here, which has resulted in flatter gradients of mono-age populations at the final time, compared to the observations (see \citealt{anders17b}). 

(3) Our finding that the kinematically hottest stars located close to the Sun today have been born locally or in the outer disk is consistent with thick disk formation from the nested flares of mono-age populations \citep{minchev15}. This phenomenon can be also seen in the observed age-velocity dispersion relation (Fig.~\ref{fig:avr}, left), in that its upper boundary is dominated by stars born at larger radii, while stars born in the inner disk are consistently cooler for most ages. A jump in the AVR around $\rm age=10$~Gyr is seen, especially well-defined for mono-$r_{birth}$ populations, which is most likely related to the Milky Way last massive merger.

(4) Our results suggest that the puzzling flatness of the local AMR is the result of the superposition of the AMRs of mono-$r_{birth}$ populations, each with a well-defined negative slope (Fig.~\ref{fig:amr}). This behavior is very similar to the predictions of classical chemical evolution models and signifies the negative ISM metallicity gradient time evolution. 

(5) We find that stars composing the metal-rich tail of the [Fe/H] distribution were born at $r\lesssim4$~kpc, consistent with angular momentum redistribution caused by the overlap of the Galactic bar and the inner spiral \citep{quillen11, minchev12a}.

(6) Based on its [Fe/H] and age, we estimated that the Sun was born at $7.3\pm0.6$~kpc, with the uncertainty being dominated by the systematics in [FeH] \citep{asplund09}.

In the next several years, with the advent of Gaia and follow up surveys, we will have at our disposal exquisite data including ages for millions of stars, covering large Galactic disk regions. This will make it possible to improve tremendously both our estimate of birth radii and the evolution of the chemical gradients with cosmic time, paving the way to a new type of Milky Way modeling, that promises to be transformative to the field of Galactic Archaeology.

\section*{Acknowledgments}
We thank the anonymous referee for thoughtful comments. 
\noindent IM acknowledges support by the Deutsche Forschungsgemeinschaft under the grant MI 2009/1-1, as well as support by Observatoire de la C\^ote d'Azur and Universit\'e C\^ote d'Azur for a one-month visit in 2017, where this project was conceived. 
ARB and PdL acknowledge financial support form the ANR 14-CE33-014-01.
CC acknowledges support from DFG Grant CH1188/2-1 and from the ChETEC COST Action (CA16117), supported by COST (European Cooperation in Science and Technology).
VA  acknowledges the supported by Funda\c{c}\~ao para a Ci\^encia e Tecnologia (FCT) through national funds (ref. PTDC/FIS-AST/7073/2014 and ref. PTDC/FIS-AST/1526/2014) and by FEDER through COMPETE2020 (ref. POCI-01-0145-FEDER-016880 and ref. POCI-01-0145-FEDER-016886). V.A. also acknowledge the support from FCT through Investigador FCT contract of reference IF/00650/2015/CP1273/CT0001.
GC acknowledges financial support from the European Union Horizon 2020 research and innovation programme under the Marie Marie Sk\l odowska-Curie grant agreement No. 664931.

\bibliographystyle{mnras}
\bibliography{myreferences}

\end{document}